# Yet another Odd Radio Circle?

Amitesh Omar[1][2]

**Abstract**

The Odd Radio Circles are newly identified diffuse radio sources at ~1 GHz frequency, with edge-brightened nearly circular morphology, which is remarkably similar to supernova remnants although a physical association with previous population of Galactic supernova remnants is challenging due to detections of the Odd Radio Circles at high Galactic latitudes. Here, a serendipitous identification of a new source in a LOFAR 144 MHz image with similar morphology as that of Odd Radio Circles is reported. This is the first reported identification of an Odd Radio Circle at a very low frequency and with the LOFAR.

---

[1] Corresponding author aomar@aries.res.in
[2] Aryabhatta Research Institute of observational sciences, Manora Peak, Nainital, 263001, India

# 1. Introduction

A new population of diffuse radio sources termed as Odd Radio Circles (ORCs) was serendipitously identified first in the deep images taken from a pilot survey using Australian Square Kilometer Array Pathfinder (ASKAP) at ~1 GHz frequency and later also in one GMRT image at 325 MHz (Norris et al. 2021). The angular sizes of these sources are fairly uniform at around 1 arcmin. The ORCs are edge brightened with a remarkable similarity with the radio morphologies of Galactic supernova remnants (SNRs). Recently, a polarization study of one ORC using the MeerKAT radio telescope revealed tangential magnetic field along the ring periphery, suggesting the ring as an expanding spherical shock wave similar to that in the expanding supernova remnants (Norris et al. 2022). The radio surface brightness of ORCs is typically in the range $10^{-21} - 10^{-22}$ W m$^{-2}$ Hz$^{-1}$ sr$^{-1}$. The 1 GHz integrated flux density of ORCs is in the range ~2 to 7 mJy.

Presently six sources have been identified with morphologies corresponding to those of ORCs. The properties of five ORCs are described in Norris et al. (2021). The geometrical centers of three ORCs seem to have a galaxy as the optical counterpart at photometric redshifts between 0.27 and 0.55. The optical colors of these galaxies suggest that the galaxies are quiescent and presently not forming stars. If ORCs are actually associated with these galaxies, the physical sizes of ORCs will be several hundred kilo-parsecs. Two ORCs are without optical counterparts and separated by 1.5 arcmin from each other. The sixth ORC was recently identified as a supernova remnant in the outskirts of the Large Magellanic Cloud (Filipovic et al. 2022). Due to their detections at high Galactic latitudes (above or near +/-40 degree) with an estimated density of 1.6 sources per 100 square degrees, ORCs cannot belong to the classical population of Galactic SNRs, which is known to be strongly concentrated near the Galactic plane.

The astrophysical origins of the ORCs are currently debated. Norris et al. (2022) considered various possibilities for their origins such as mergers of black holes, winds from starburst galaxies and end-on lobes of radio galaxies. Omar (2022a) showed that some ORCs can be SNRs in the intragroup medium associated with the immediate neighbor groups of galaxies, and hence can be detected across the sky. Omar (2022b) presented another possibility that ORCs may be created during a burst of tidal disruption events in post starburst galaxies, where the required energies for a large radio halo can come from the ejected unbound debris, creating energetic shocks in the interstellar and intergalactic medium, similar to those created via the supernova remnants in a starburst like conditions in a galaxy.

## 2. Odd Radio Circle in LOFAR 144 MHz Image

The author was working with some three odd images downloaded from the Low Frequency Array (LOFAR) Two-meter Sky Survey (LoTSS) database (Shimwell et al. 2022) at center frequency of 144 MHz in the band of 122 MHz — 168 MHz. The LoTSS field identifications and their approximate Galactic coordinates (latitude 'l' and longitude 'b' in degrees) are P160+35 (l=190, b=61), P000+23 (l=108, b=-38), and P006+31 (l=115, b=-32). Each image covers an area of ~10 square degrees. These images were being studied for a project on identifying morphologies of some radio galaxies and hence were randomly selected. The LoTSS images show a plethora of diffuse radio morphologies, many of those being identified for the first time and hence the author explored full extents of each LoTSS image just out of curiosity. The author has examined only three LoTSS images so far. A serendipitous identification of a circular object similar to the previously published ORCs was made in one field (P006+31). The location of the ORC is far from our region of interest in the field. Figure 1 shows LoTSS radio images and the corresponding optical image in the *i*-band taken from the Panoramic Survey Telescope and Rapid Response System (Pan-STARRS; Flewelling et al. 2020) database. The Galactic coordinates of this source are l=115.1 degree and b=-32.1 degree. The angular size is nearly 1 arcmin. The integrated 144 MHz radio flux is estimated to be ~32 mJy. For an assumed spectral index $\alpha = -1$ ($S \sim \nu^{\alpha}$), the predicted flux at 1 GHz is ~4.6 mJy.

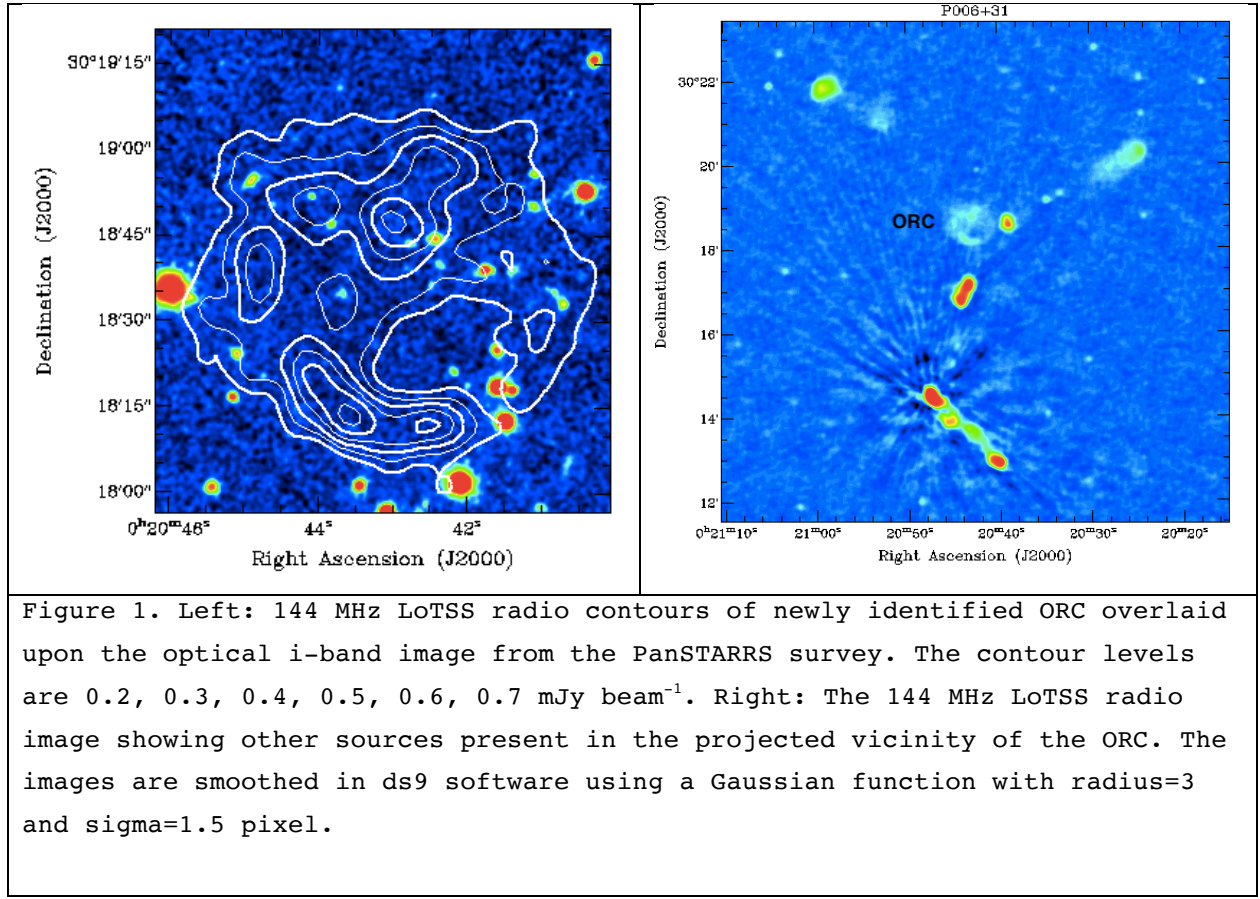

Figure 1. Left: 144 MHz LoTSS radio contours of newly identified ORC overlaid upon the optical i-band image from the PanSTARRS survey. The contour levels are 0.2, 0.3, 0.4, 0.5, 0.6, 0.7 mJy beam$^{-1}$. Right: The 144 MHz LoTSS radio image showing other sources present in the projected vicinity of the ORC. The images are smoothed in ds9 software using a Gaussian function with radius=3 and sigma=1.5 pixel.

This ORC has edge-brightened morphology and does not appear similar to normal morphologies of radio galaxies or newly identified rare morphologies of radio sources in the LoTSS images from a machine learning algorithm (Mostert et al. 2021). There are a number of radio galaxies with classical lobes or jetted radio morphologies in the vicinity of this source, although the ORC cannot be readily considered as an extension of any of these sources. This newly identified ORC does not seem to have an obvious optical counterpart also, although there are a number of optical sources within its extent. A previous case of ORC without an optical counterpart also has a nearby radio galaxy (see ORC-2 in Norris et al. 2021).

**3. Concluding Remarks**

The newly identified nearly circular edge-brightened radio source as reported in this note is the first one from LOFAR and at the lowest frequency of 144 MHz. The morphology of this source is similar to the recently discovered ORCs in the ASKAP survey at ~1 GHz and the GMRT image at 325 MHz. The angular size and radio flux are also similar to the previously known ORCs. When compared to some other known ORCs, this ORC also appears to be in an optically over-dense region although no optical galaxy could be identified within the extent of the ORC.

If we consider that finding of one source in three random LoTSS field images, each ~10 degrees square, reflects the general population density then a density of 1 ORC per ~30 square degrees may be possible in the LoTSS images. This detection rate appears suitable for various citizen scientists to explore LoTSS images and identify new ORCs. It is then somewhat surprising that no ORC-like object has been published so far from the LoTSS images, including in a search made using an automated machine learning algorithm (Mostert et al. 2021). The reasons for that could be — (i) presently used machine learning algorithms are not sufficiently trained for detecting edge-brightened circular objects on all scales, (ii) researchers using LoTSS images are not examining the full extents of images outside the regions of interest, and (iii) ORCs have already been detected by some research group using a targeted machine learning search but results are not yet published. Either way, this new detection reported here should encourage researchers and citizen scientists to carefully examine LoTSS images for potential new types of radio sources including ORCs.